\newcommand{\Lis}{$^{6}$Li}
\newcommand{\Bt}{$^{10}$B}
\newcommand{\Cftft}{$^{252}$Cf}
\newcommand{\Csots}{$^{137}$Cs}

\newcommand{\Baott}{$^{133}$Ba}
\newcommand{\Amtfo}{$^{241}$Am}

\newcommand{\degree}{\ensuremath{^{\circ}}}

\newcommand{\nuc}[2]{\ensuremath{^{#1}}#2}          
\newcommand{\reaction}[6]{\nuc{#1}{#2}(#3,#4)\/\nuc{#5}{#6}}

\documentclass[5p,twocolumn]{elsarticle}
\usepackage{color}
\usepackage{lineno}
\usepackage{amssymb}

\journal{Nuclear Instruments and Methods A}

\begin{document}
\begin{frontmatter}

\title{Fast Neutron Detection with \Lis-loaded Liquid Scintillator}

\author[a]{B.~M.~Fisher}\fnref{label3} \ead{brian.fisher@jhuapl.edu}
    \address[a]{National Institute of Standards and Technology, Gaithersburg, MD 20899 USA}
\fntext[label3]{Present address: Johns Hopkins Applied Physics Laboratory, Laurel, MD, USA}
\author[b]{J.~N.~Abdurashitov}
    \address[b]{Institute for Nuclear Research, Russian Academy of Sciences, Moscow, 117312 Russia}
\author[c]{K.~J.~Coakley}
    \address[c]{National Institute of Standards and Technology, Boulder, CO 80305 USA}
\author[b]{V.~N.~Gavrin}
\author[a]{D.~M.~Gilliam}
\author[a]{J.~S.~Nico}
\author[b]{A.~A.~Shikhin}
\author[a]{A.~K.~Thompson}
\author[c]{D.~F.~Vecchia}
\author[b]{V.~E.~Yants}
\date{\today}

\begin{abstract}

We report on the development of a fast neutron detector using a liquid scintillator doped with enriched \Lis. The lithium was introduced in the form of an aqueous LiCl micro-emulsion with a di-isopropylnaphthalene-based liquid scintillator. A \Lis\ concentration of 0.15\,\% by weight was obtained.  A 125\,mL glass cell was filled with the scintillator and irradiated with fission-source neutrons. Fast neutrons may produce recoil protons in the scintillator, and those neutrons that thermalize within the detector volume can be captured on the \Lis. The energy of the neutron may be determined by the light output from recoiling protons, and the capture of the delayed thermal neutron reduces background events. In this paper, we discuss the development of this \Lis-loaded liquid scintillator, demonstrate the operation of it in a detector, and compare its efficiency and capture lifetime with Monte Carlo simulations. Data from a boron-loaded plastic scintillator were acquired for comparison. We also present a pulse-shape discrimination method for differentiating between electronic and nuclear recoil events based on the Matusita distance between a normalized observed waveform and nuclear and electronic recoil template waveforms. The details of the measurements are discussed along with specifics of the data analysis and its comparison with the Monte Carlo simulation.

\end{abstract}


\end{frontmatter}

\section{Overview}
\label{sec:overview}

Fast neutrons may be produced through several mechanisms.  Naturally occurring isotopes in the \nuc{238}{U} and \nuc{232}{Th} decay chains generate fast neutrons via spontaneous fission,  and ($\alpha$,n) reactions create neutrons in the 1\,MeV to 15\,MeV range. Particle accelerators produce higher energy fast neutrons through a variety of reactions, and neutrons of very high energies are generated from cosmic-ray muon-induced spallation reactions~\cite{DEL95}. Improvements in the ability to detect and characterize these neutrons are of interest to diverse research interests such as fundamental physics, neutron dosimetry, and detection of low-level neutron emissions. 

In the area of basic physics research, many classes of nuclear, particle, and astrophysics experiments must be performed in underground laboratories to reduce the backgrounds generated by naturally occurring radioactivity and cosmic rays~\cite{FOR04}.  Although experimenters operate their detectors deep underground and go to great lengths to optimize radiation shielding, the experiments have become so sensitive that even these efforts may not be sufficient. The characterization of the fast neutron fluence has become a critical issue for experiments that require these extreme low-background environments, such as neutrinoless double-beta decay~\cite{ELL02, AAL05, SCH06}, dark matter searches~\cite{AKE03, GAI04, BOU04, ANG05}, and solar neutrino experiments~\cite{CLE98, HAM99, FUK01, ABD09, MCK05, AHA07}.  In some experiments, fast neutrons may be the dominant and potentially irreducible background, thus necessitating precise information about the fast neutron fluence and energy spectrum.  The most reasonable approach to addressing the problem is through the complete characterization of the neutrons through both site-specific measurement~\cite{MEI06} and benchmarking of simulation codes~\cite{WUL04, MAR07}. 

The health physics community is another area where improved fast neutron detection and spectroscopy are needed~\cite{ISO99}. Existing spectrometers allow the determination of fluence spectra (and dose) for low energy neutron sources such as \Cftft\ and D$_{2}$O  moderated \Cftft, but begin to have difficulty with higher energy sources such as \Amtfo-Be($\alpha$, n) or \Amtfo-B($\alpha$, n).  Current spectrometers fail almost completely for  determining neutron fields with energies of tens of MeV and may require multiple measurements with different detectors and complicated unfolding procedures.  This need has only grown due to the increased use of 14\,MeV neutron generators in interdiction and inspection technologies. Without good knowledge of these neutron spectra, health physicists cannot accurately calibrate radiation protection instruments and dosimeters, which may result in incorrect determination of dose received by radiation workers.

An improved fast neutron detector has direct application to the detection of low fluence rates of fast neutrons, such as from fissile material.  The technological challenges and requirements for measuring the fast neutron fluence in the underground environment are very similar to this problem.  A highly efficient neutron spectrometer with reasonable resolution would be capable of detecting low-level neutron signals from a number of sources. 

In this paper, we present the developmental work performed using a small detector filled with an enriched \Lis\ liquid scintillator. General methods of neutron detection and spectroscopy may be found elsewhere~\cite{KNO89,BRO02}. Section~\ref{sec:ndetection} discusses capture-gated neutron detection and the fabrication of the \Lis-loaded liquid scintillator. In Section~\ref{sec:det} the results of measurements with a test detector of \Lis-loaded liquid scintillator and a B-loaded plastic scintillator are presented along the analysis methods. Section~\ref{sec:psd} presents a pulse-shape discrimination method for differentiating between electronic and nuclear recoil events.

\section{Neutron Detection with \Lis\ Scintillator}
\label{sec:ndetection}

\subsection{Capture-Gated Neutron Detection}
\label{subsec:intro:capturegated}

\begin{figure}
    \includegraphics[width=\columnwidth]{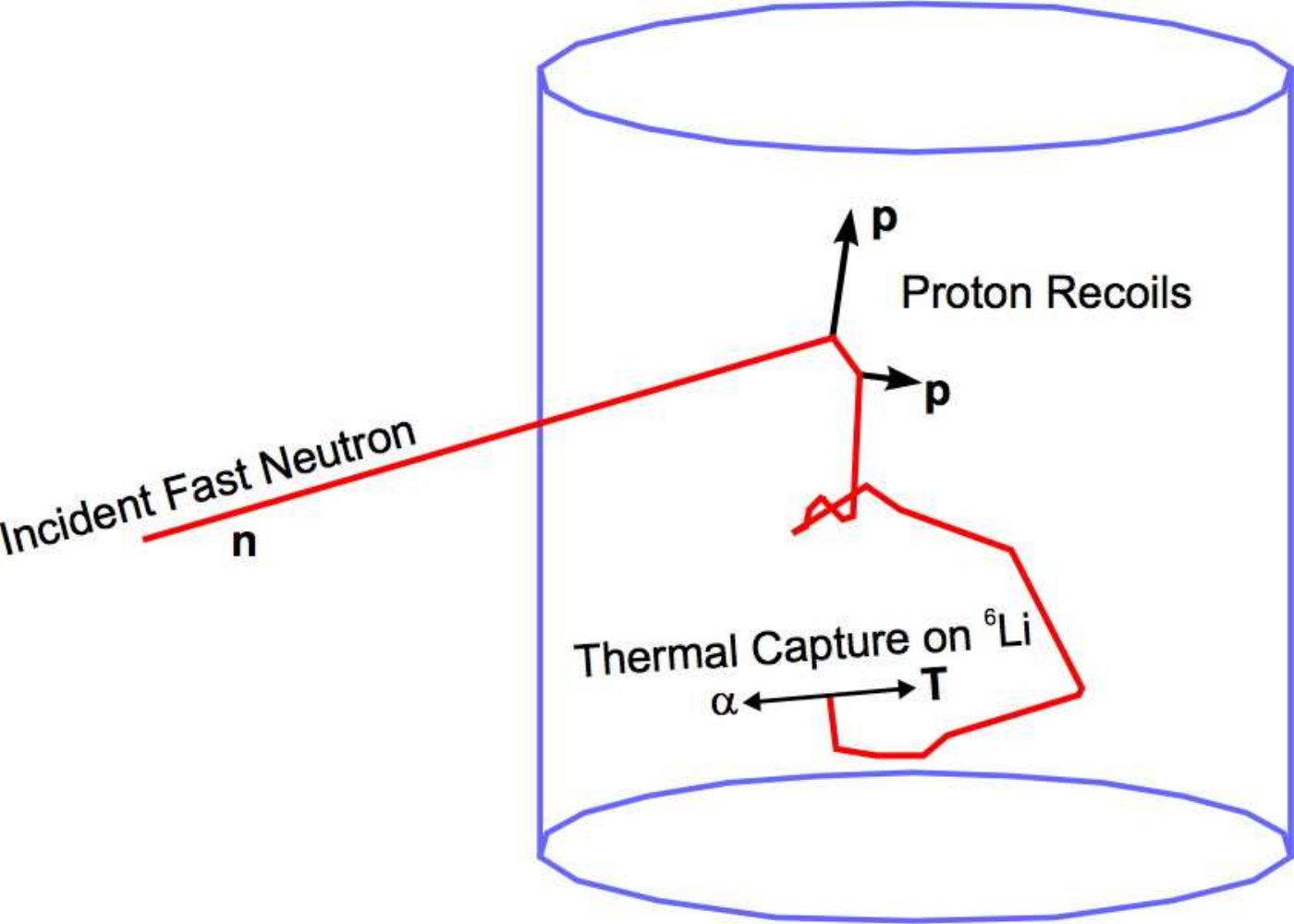}
        \caption{An illustration of the principle of capture-gated detection. A fast neutron impinges on the detector. It rapidly gives up its energy through nuclear collisions, primarily with protons, in the moderation process. The thermalized neutron diffuses in the medium until it is captured on a material with a high capture cross section.}
    \label{fig:cap_gate}%
\end{figure}

The detection method used in this work is known as capture-gated neutron spectroscopy~\cite{DRA86, CZI89, CZI02, NOR02}. As illustrated in Figure~\ref{fig:cap_gate},  an incident fast neutron preferentially scatters from a proton in a scintillation medium. The proton recoils with approximately half (on average) of the neutron's energy, producing scintillation light.  The neutron then scatters off another proton, producing another proton recoil, and continues to scatter until it leaves the detector volume or is captured, typically at thermal energies. A large fraction of the neutrons that lose all their energy within the scintillator volume and then capture will have lost most of their energy through neutron-proton scattering, and the majority of the detected light will come from those neutron-proton events. There are other mechanisms that may generate light, in particular scattering from carbon~\cite{POZ07}, but these are small effects for the measurements performed here.

An organic scintillator provides a high density of protons that will efficiently moderate the neutron; a neutron with an energy of a few MeV loses 90\,\% of its energy in this manner during the first 10~ns in the scintillator. The thermalized neutron diffuses within the bulk of the scintillator and may capture on an isotope that was introduced because of its large neutron capture cross section and unambiguous signature of the neutron capture. The time to capture occurs within approximately 100~${\rm{\mu}}$s and depends upon upon the size and geometry of the detector and the concentration and neutron capture cross section of the isotope used as the dopant. The delayed-coincidence signature indicates that the fast neutron gave up essentially all of its energy within the scintillator and provides a method of rejecting of uncorrelated backgrounds.
 
One must consider several factors when selecting which capture isotope is appropriate for the application. Typical elements are gadolinium, boron, and lithium. Gadolinium is commercially available and has a very large neutron capture cross section, but it produces gamma rays of various energies (up to 9\,MeV) that may be difficult to detect with small volume detectors. In addition, the Gd gamma rays cannot be distinguished from background gamma rays by means of pulse shape discrimination. With neutron capture on \Bt, the \nuc{7}{Li} can be formed in a ground or excited state, leading to two reaction branches:

\begin{eqnarray}
^{10}{\rm B} + n  \rightarrow {^{7}{\rm Li}}^*\,(0.84\,{\rm MeV})+ \alpha\, (1.47\,{\rm MeV}) \\ \nonumber
^{7}{\rm Li}^* \rightarrow ^{7}{\rm Li} + \gamma\,(0.477\,{\rm MeV}) \\
^{10}{\rm B} + n  \rightarrow {^{7}{\rm Li}}\,(1.01\,{\rm MeV})+ \alpha\, (1.78\,{\rm MeV}).
\end{eqnarray}

\noindent The branching ratio for the first reaction is 93.7\,\% and 6.3\,\% for the second. The cross section is also large, and the alpha particle makes efficient thermal detection possible with a small-volume detector. Boron-loaded detectors are expensive, which tends to make scaling to larger-volume detectors cost-prohibitive.

A good alternative to \Bt\ is \Lis\ 

\begin{equation}
 ^{6}{\rm Li} + n \rightarrow t\,(2.05\,{\rm MeV}) + \alpha\,(2.73\,{\rm MeV}).
\end{equation}

\noindent The use of enriched \Lis\ as the dopant has the advantage of a large Q-value and the production of two energetic charged-particles. The light yield of the 2-MeV triton is nearly a factor of 10 higher than that of the 1.5-MeV alpha from neutron capture on \Bt. The neutron capture signature from \Lis\ is well-separated from the noise and most background sources. There is no concern about the energy leaving the scintillator as there is with dopants that produce gamma rays, and pulse shape discrimination between gammas and the capture products is possible. The capture cross section is high, but it less than for Gd and \Bt.  

The capture-gated detection method using \Lis\ has been implemented in different ways. A 5-L detector was constructed for measurements in the Gran Sasso Laboratory~\cite{ALE89}, and an 8-L detector was used to measure the fast neutron background at the Modane Underground Laboratory in Modane, France~\cite{CHA98}.  Both detectors used the commercial \Lis-doped liquid scintillator NE320~\cite{AIT89}. In a different approach, a plastic scintillator impregnated with lithium gadolinium borate crystals was developed and was demonstrated to detect neutrons over a wide range of energies~\cite{CZI02}. Studies of the response of B-loaded liquid scintillator to monoenergetic neutrons have also been carried out~\cite{AOY93}.

\subsection{\Lis-loaded Liquid Scintillator}
\label{sec:scintillator}

\begin{table}
\begin{center}
\begin{tabular}{lccccc}
\hline
 Isotope	 & Scintillator	&	$\sigma_n$	&$f$(\%)	& Volume		& H/C  \\
			\hline
\Lis\	 	&	liquid 	&	941\,b			&	0.15		&	125 cm$^3$		&	1.5 \\
B	 	&	plastic	&	3842	\,b			&	1.1		&	103 cm$^3$		&	1.1 \\
\hline
\end{tabular}
\end{center}
\caption{\label{tab:scints} Properties of the two scintillator materials used in testing. $\sigma_n$ is the 2200\,m/s neutron capture cross section; $f$ is the fractional weight of the isotope in the scintillator; and H/C is the hydrogen to carbon ratio of the scintillator.}
\end{table}

Natural lithium for use as a neutron capture material was introduced into organic liquid scintillators more than 50 years ago~\cite{KAL56, HEJ61}.  More recently a solution made using \Lis-salicylate dissolved in a toluene-methanol solvent~\cite{GRE79} was developed.  Scintillators with enriched \Lis\ were developed later, and in the 1980s Nuclear Enterprises manufactured a lithium-loaded pseudocumene-based scintillator (NE320~\cite{TRA08}), which was used to make several measurements of neutron backgrounds in underground laboratories \cite{AIT89, ALE89, CHA98}.  This scintillator is no longer being manufactured, and to the authors' knowledge there are currently no  vendors of scintillator doped with enriched \Lis.  Hence, detectors incorporating \Lis\ have not been made for many years due to the lack of commercial availability of a scintillator.

Given the possible applications for a \Lis-doped scintillator, we have investigated approaches for introducing lithium compounds into liquid scintillators.  After exploring methods for incorporating organolithium compounds directly into an organic scintillator, we investigated technology used in liquid-scintillation counting in the life sciences.  Radioactive samples used in biochemistry are often low-energy beta emitters in an aqueous solution, such as tritium in urine.  The very short range of these low-energy $\beta$-particles make traditional counting very difficult.  The liquid scintillators used in this field have additional surfactants that allow for the creation of micro-emulsions between the aqueous solution and the organic liquid scintillator.  This permits the counting sample to be uniformly distributed throughout the scintillator, greatly increasing its counting efficiency for low energy $\beta$-particles.

A critical property of the desired organic scintillator was its ability to distribute and suspend a lithium-containing compound uniformly throughout its volume. We found a scintillator manufactured by Zinsser Analytic~\cite{TRA08} composed of a solvent (very high-purity di-isopropylnaphthalene) with added surfactants allowing introduction of aqueous solutions up to 40\,\% water by volume.  Di-isopropylnaphthalene is considered a safer solvent than the more common pseudocumene-based liquid scintillators. It has a higher flashpoint (approximately 150\,{\degree}C) and is biodegradable, allowing for safer handling and operation.

\begin{figure}[t]
    \includegraphics[width=\columnwidth]{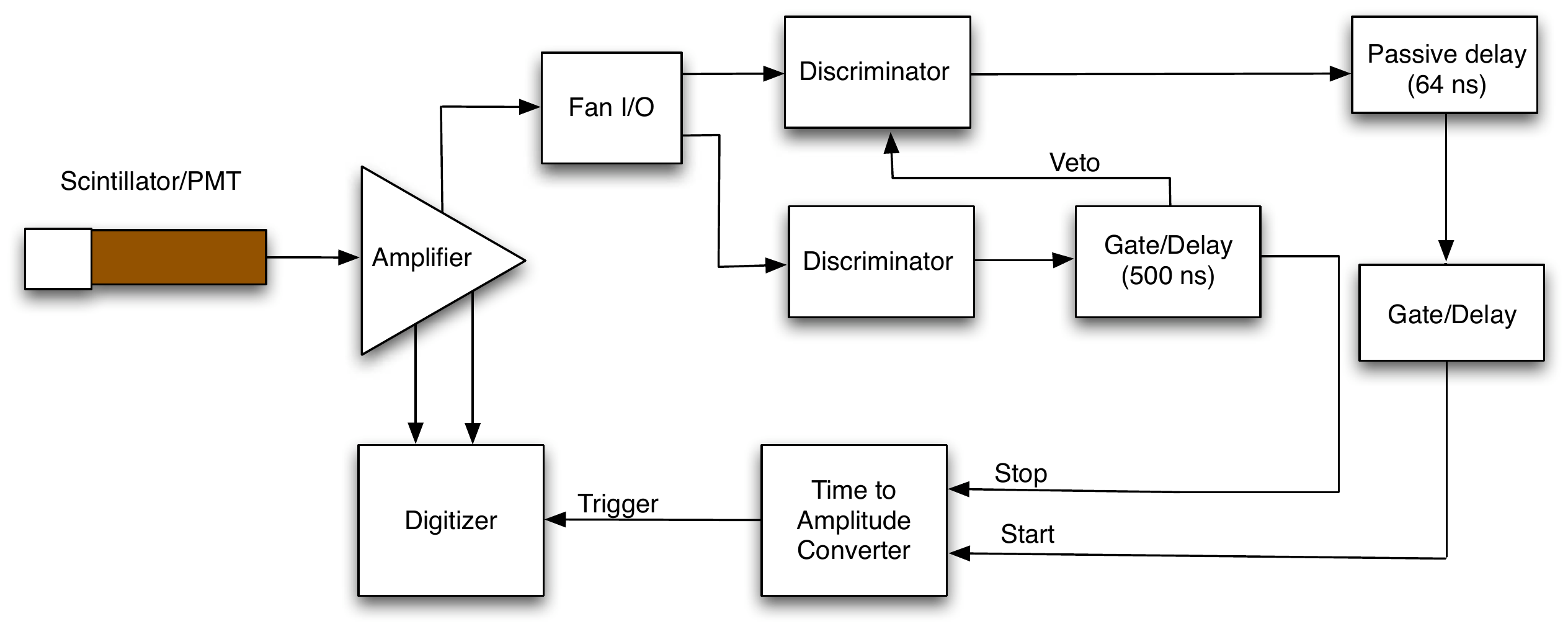}
        \caption{Block diagram of the data acquisition electronics. The two inputs to the digitizer allow for different range settings.}
    \label{fig:electronics}
\end{figure}

A 10-molar aqueous solution of \Lis-enriched LiCl was created by reacting enriched Li$_{2}$CO$_{3}$ (95\,\% \Lis\ by weight) with concentrated hydrochloric acid, boiling off the excess acid, and dissolving the remaining dried LiCl salt in pure de-ionized water.  Mixing this LiCl (aqueous) solution with the liquid scintillator, we achieved \Lis\ concentrations of approximately 0.15\,\% by weight~\cite{FIS04}.  This concentration is comparable to what had been commercially available and used previously~\cite{CHA98, ALE89}.  This scintillator mixture was found to have good light properties, and no precipitation of LiCl was seen during the duration of testing.

\section{Detector Characterization}
\label{sec:det}

\subsection{\Lis-loaded Liquid Scintillator Detector}
\label{subsec:det:fastneutron}

A 125-mL cell was filled with this scintillator and inserted into fast neutron fields to demonstrate its effectiveness as a neutron detector. Some characteristics of the cell and scintillator are given in Table~\ref{tab:scints}. A 5.08-cm diameter photomultiplier tube (PMT, Burle 8850) was mounted on one face of the 5.08-cm diameter by 5.08-cm long cylindrical cell, and the remaining walls were coated with a diffuse, white reflector. The logic of the data acquisition electronics is shown in the block diagram of Fig.~\ref{fig:electronics}. The PMT signal was amplified and split into two signals. One output was delayed 64\,ns to avoid self-triggering, and it became the start signal in a time-to-amplitude converter. To further reduce the probability of self-triggering, the discriminator was vetoed for 500\,ns after an event.  The trigger was formed on any event above a threshold followed by a second event that occurred within 40\,$\mu$s; the waveform containing both events was digitized and recorded to disk. The data acquisition system was a single 2-channel 2\,GHz digital oscilloscope card with 8-bit resolution.

\begin{figure}[t]
    \includegraphics[width=\columnwidth]{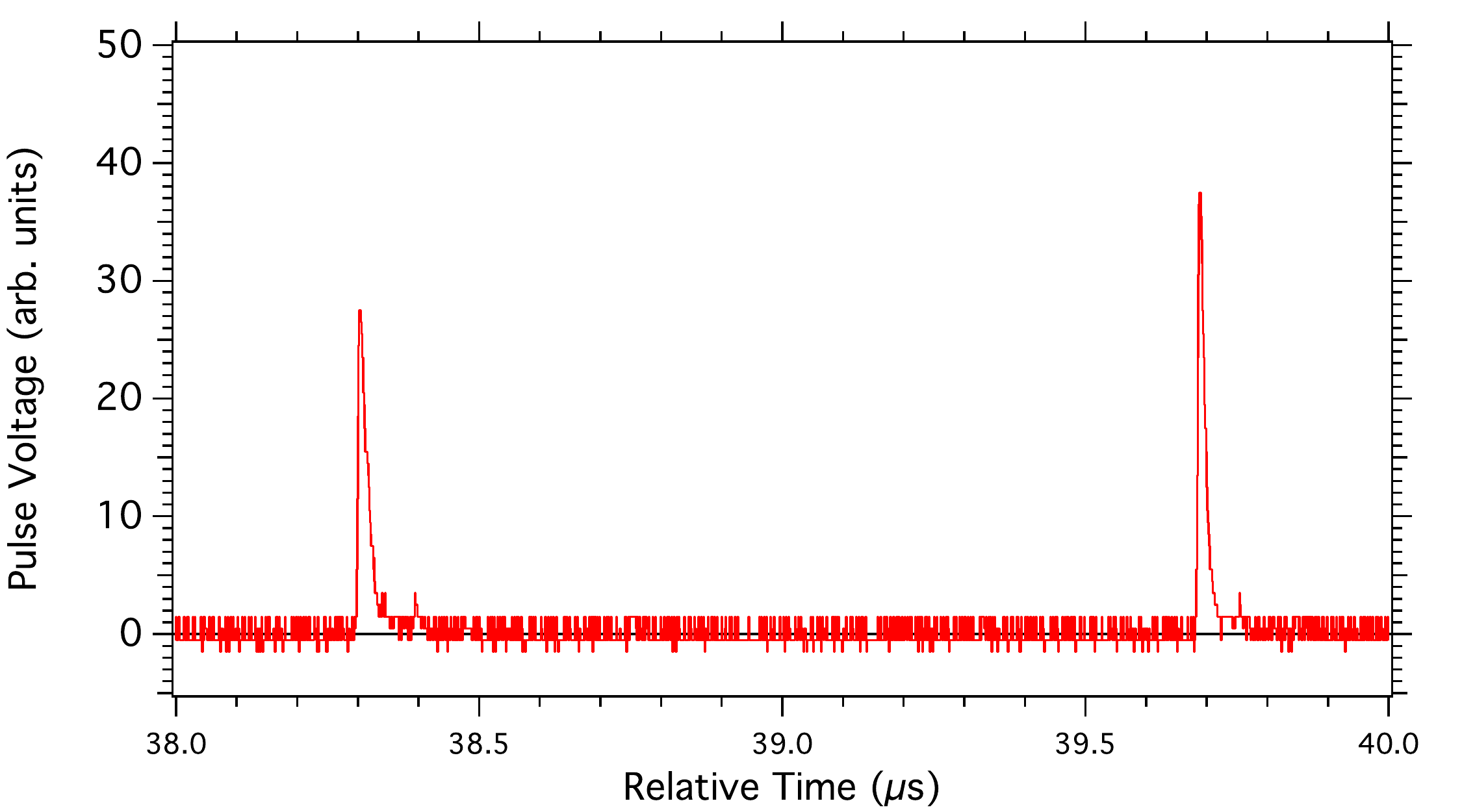}
        \caption{Plot of the digitized PMT pulse of the detector response to a fast neutron within the scintillator. A neutron enters the test cell and initially scatters off the protons in the scintillator.  After thermalizing within the scintillator, it captures on a \Lis, in this example approximately 1.4\,$\mu$s later.}
    \label{fig:waveform}
\end{figure}

Figure~\ref{fig:waveform} gives a sample trace showing the initial proton recoil of a fast-neutron off hydrogen in the scintillator followed by the \Lis-capture pulse. All of the analysis was performed on these digitized waveforms. Each waveform was analyzed to determine the energy of each pulse and the relative timing of the two events. The pulse offset was determined by averaging 200\,ns of the baseline near the stop event. The pulse height corresponds to the maximum voltage of the pulse, and the time of the pulse was obtained from the timing channel corresponding to the position of the maximum pulse height.  Histograms of these parameters from the measured data were constructed from the analysis of all events.

The light yield of the scintillator is nonlinear as a function of proton recoil energy. Thus,  the sum of all the light from individual recoil events is not proportional to the incident neutron energy. This results in the degradation of the energy resolution. By segmenting the detector, one can ensure that the probability of having more than one energy deposit in any segment is very small and thus reconstruct the initial neutron energy by summing the light output of the individual segments~\cite{ABD02b}.  This produces a detection scheme that can still be efficient while achieving good energy resolution and suppression of uncorrelated backgrounds. There is an effort in progress to construct a 16-channel pilot spectrometer based on this idea~\cite{ABD07}.

\begin{figure}[t]
    \centering
    \includegraphics[width=\columnwidth]{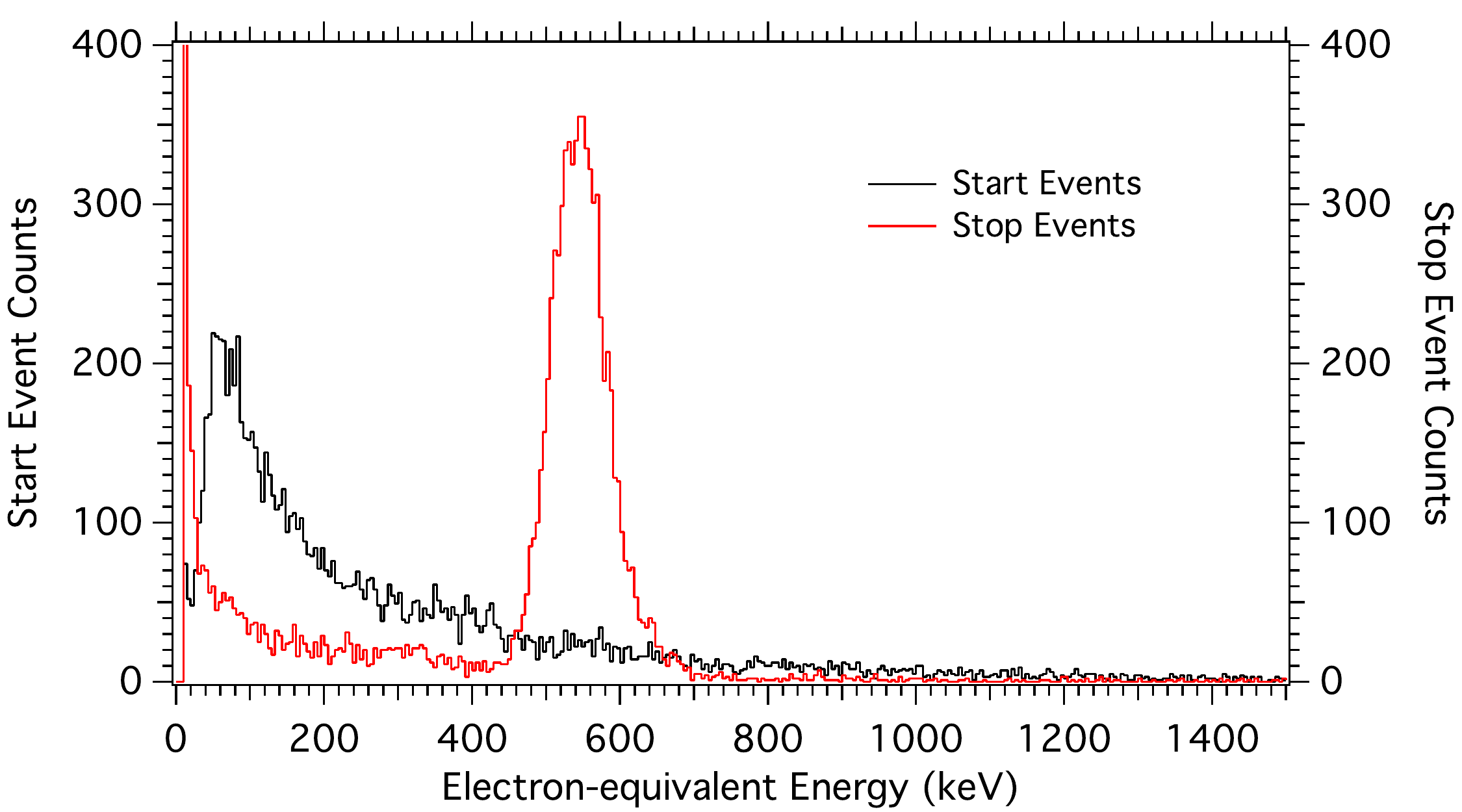}
    \includegraphics[width=\columnwidth]{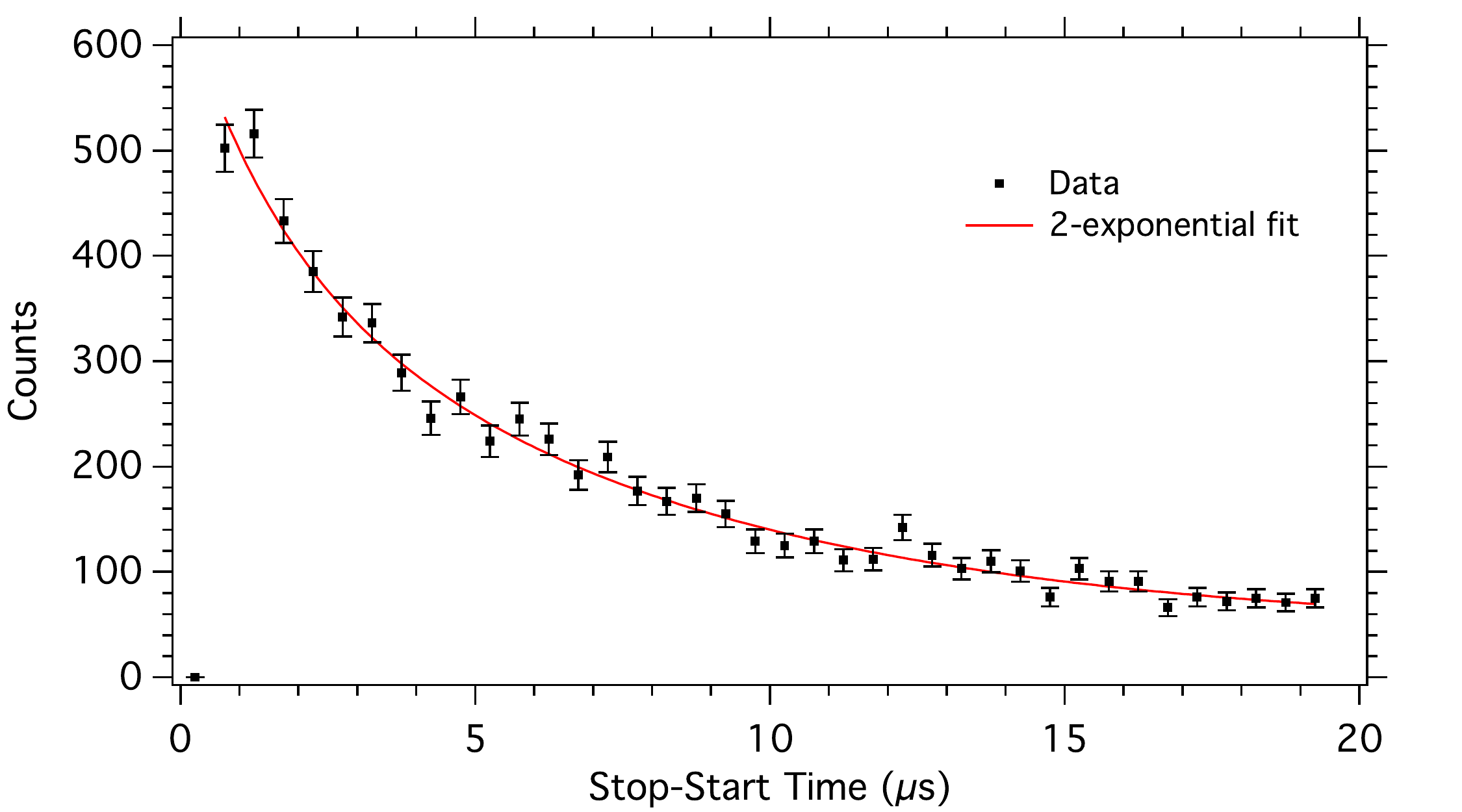}
    \caption{Results from a demonstration of the \Lis-loaded liquid scintillator cell irradiated by a \Cftft\ source. The top plot shows the energy distribution of the start events (primarily recoil events) and stop events (primarily capture on \Lis). The peak in the spectrum of stop events corresponds to the total energy of the \Lis\ reaction products.  The bottom plot shows the delay time (between start and stop events) distribution for event sequences where the stop event produces light consistent with neutron capture on \Lis. The error bars represent plus and minus the combined standard uncertainty.}
    \label{fig:XLSdemo}
\end{figure}

The test cell was exposed to fission neutrons from \Cftft\ sources and 2.5-MeV monoenergetic neutrons from a commercial neutron generator.  Both the sources and generator are maintained at the NIST Californium Neutron Irradiation Facility (CNIF)~\cite{GRU77}.  The top plot of Figure~\ref{fig:XLSdemo} shows the energy spectrum for both the start and stop pulses of the trigger when the detector was irradiated with neutrons from a \Cftft\ source. The start events correspond primarily to proton recoil events while the stop events primarily correspond to the capture of the thermalized neutron on \Lis. The peak in the capture spectrum is due to the reaction products from the \reaction{6}{Li}{n}{$\alpha$}{3}{H} reaction; most of the scintillation light comes from the triton. The peak is clearly separated from the noise, and the signal-to-background is more than 10:1. The energy spectra were calibrated using the Compton edges of gamma sources such as \Baott\ and \Csots, as there is little photopeak detection in the liquid scintillator. Note that the energy scale for both plots is given in the electron-equivalent energy. The timing data were fit to a two-component exponential with a constant background, as shown in the bottom plot of Figure~\ref{fig:XLSdemo}. The error bars represent plus and minus the combined standard uncertainty. A long component arises from the capture of thermalized neutrons as they diffuse within the bulk of the scintillator.  The simulation indicates that at times below $1\,\mu$s there is a population of partially thermalized neutrons that can still capture on \Lis\ or \Bt. These events give rise to a second, faster component in the timing spectrum.

Because the data were digitized event-by-event, recoil events that do not correspond to a valid neutron capture can be rejected in analysis. Two important cuts are the energy of neutron capture and the timing of the capture. An energy cut can be made around the \Lis\ capture peak (shown in Fig.~\ref{fig:XLSdemo}) to reject with high probability recoil events that do not have a valid capture associated with them. The capture peak was fit to a gaussian function, and the energy window was chosen to be $\pm 3$ sigma around the peak channel of the distribution. The bottom plot of Fig.~\ref{fig:XLSdemo} shows the time distribution of capture times for the thermalizing neutron after the energy cut was made to keep only events with neutron captures on \Lis. One can use the timing spectrum to perform another cut. At long times, the correlated capture events are gone, and one is left with the uncorrelated background. The spectrum of these background events can be subtracted from energy spectrum of the recoil events. Typically, the signal window was $0.5\,\mu$s to $10\,\mu$s and the background window was $10\,\mu$s to $20\,\mu$s. The relative size of the windows could vary slightly from run to run depending on the signal-to-background in the particular time spectrum.

Data were also acquired in a similar data manner using plastic scintillator loaded with 5\,\% natural boron (BC-454), corresponding to about 1\,\% \Bt. Some its characteristics are given in Table~\ref{tab:scints}. The scintillator was a 5.08\,cm by 5.08\,cm right cylinder coupled to a 5.08\,cm diameter PMT (Burle 8850).  The setup, data acquisition, and analysis were carried out in the same manner as that done for the \Lis-loaded cell. These data were useful for comparison with the \Lis\ data and also as another check on the computer simulations. 

\begin{figure}[t]
    \centering
    \includegraphics[width=\columnwidth]{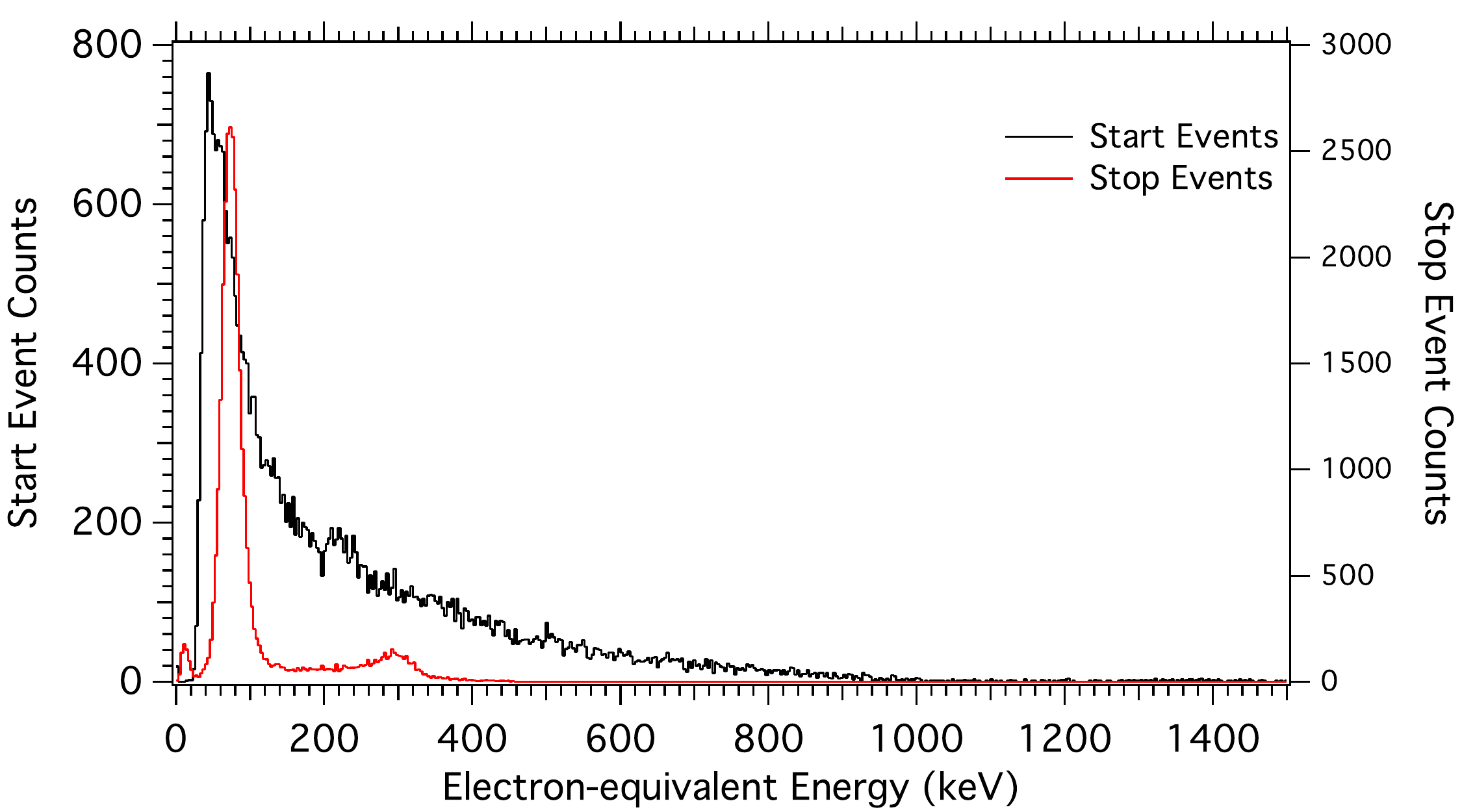}
    \includegraphics[width=\columnwidth]{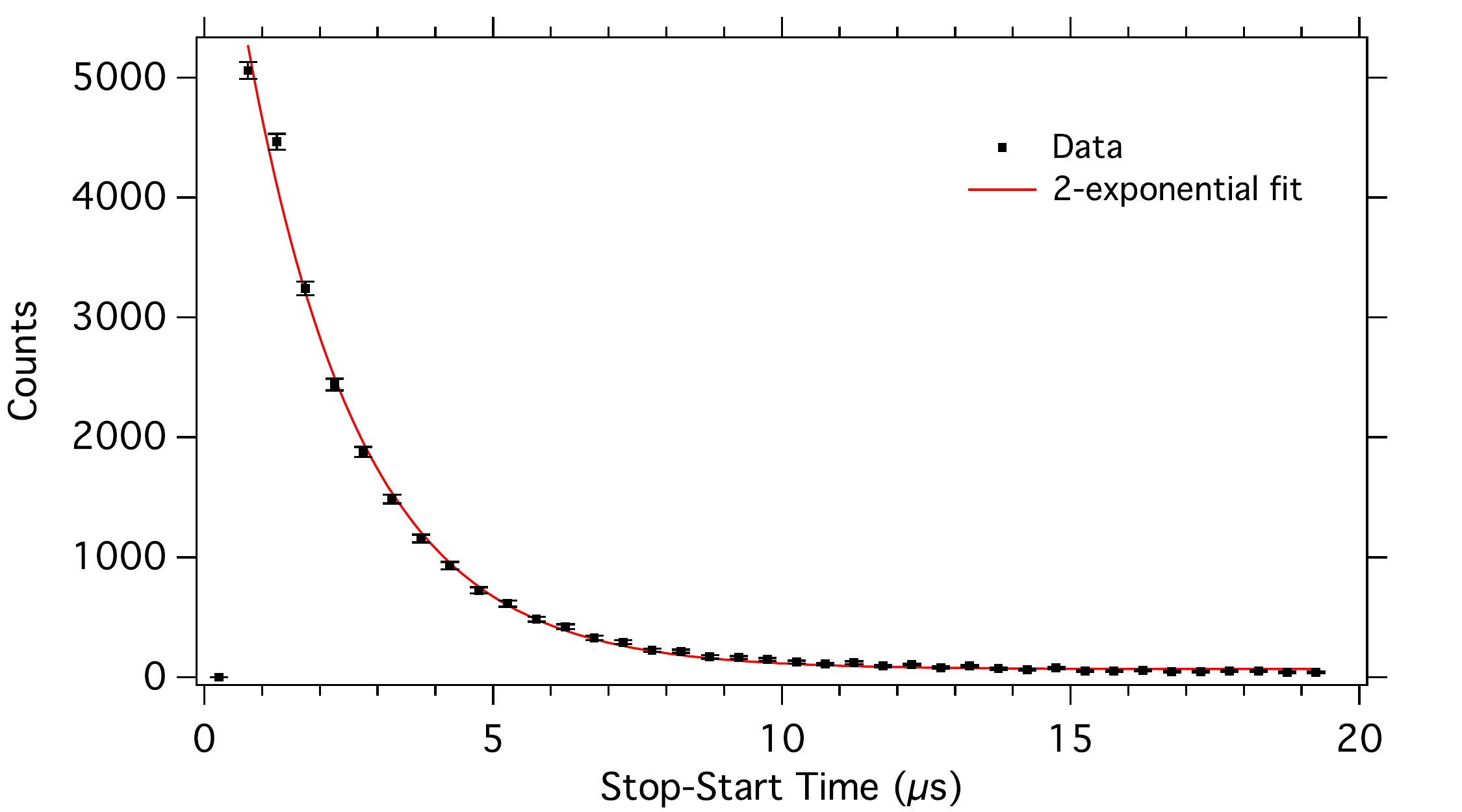}
    \caption{The top plot shows the energy distribution of the start and stop events for the irradiation of plastic scintillator by neutron from \Cftft. The large peak in the spectrum of stop events comes from the alpha particle and the smaller peak to the right is the Compton edge of the 477\,keV gamma. The bottom plot shows the time distribution of all events that have a valid neutron capture on \Bt.}
    \label{fig:B10demo}
\end{figure}

The top plot of Figure~\ref{fig:B10demo} shows the energy spectrum for both the start and stop events when the detector was irradiated with neutrons 
from a \Cftft\  source. The energy scale was calibrated using a \Baott\ source and the Compton edge of the 477\,keV photon, which is seen in the stop spectrum. The number of events for the gamma is less than for the alphas due to the lower efficiency for stopping the gammas in the small volume of plastic. The bottom plot of Figure~\ref{fig:B10demo} shows the time distribution of neutron captures on the \Bt. As with the \Lis\ cell, the distribution was made after an energy cut on alpha peak in the stop spectrum, thus requiring that there was a valid neutron capture. The distribution of delayed-capture times fits well to the 2-component exponential function.

\begin{table*}
\begin{center}
\begin{tabular}{|l|ccccccc|}
\hline
			& Pb shield	&	Activity				& Number of 		&	Range		& 	$R_{Tot}$		&	$R_{bkgd}$	& Threshold	\\
			&			&	(Bq)					& Positions 		&	(cm)			& 	(s$^{-1}$)		&	(s$^{-1}$)		& (keV)		\\
\hline
 \Lis\ liquid 	&	off		&	$1.2\times10^6$		&	7			&	107 -- 230		&	3.90			&	3.77			&	75		\\
scintillator		&	on		& 	$1.2\times10^6$		&	6			&	86 -- 230		&	3.72			&	3.40			&	76		\\
			&	off		& 	$1.0\times10^3$		&	6			&	8 -- 33		&	0.308		&	0.258		&	86		\\
			&	on		& 	$1.0\times10^3$		&	6			&	9 -- 30		&	0.0530		&	0.023		&	93		\\
			&			&						&				&				&				&				&			\\

B-plastic	 	&	off		&	$1.2\times10^6$		&	7			&	81 -- 231		&	5.07			&	4.71			&	29		\\
scintillator		&	on		& 	$1.2\times10^6$		&	7			&	74 -- 221		&	3.32			&	2.01			&	31		\\
			&	off		& 	$1.0\times10^3$		&	5			&	8 -- 30		&	0.729		&	0.577		&	30		\\
			&	on		& 	$1.0\times10^3$		&	5			&	8 -- 30		&	0.149		&	0.021		&	32		\\
\hline
\end{tabular}
\end{center}
\caption{\label{tab:params} Table of some of the run parameters for the  \Lis\ liquid scintillator and the B-plastic scintillator. The first column states whether the scintillator had a lead shield; the second column indicates the activity of the \Cftft\ source; the third column gives the number of runs that were done at different positions; the fourth column gives the range of source-detector distances $r$; the fifth column gives the measured signal rate (at the closest position only); the sixth column gives the background rate (at the closest position only); and the seventh column list the low-energy threshold that was used for both the data and simulation.}
\end{table*}

\subsection{Efficiency Measurements}
\label{subsec:efficiency}

Both the \Lis\  cell and the B-plastic scintillator were irradiated by \Cftft\ sources of known activity in the CNIF, and measurements of their efficiencies for fast neutron detection were performed. Irradiations  were carried out using two source activities and two gamma shielding configurations, and the measurements were done at several source-detector distances. For the gamma shielding, there was either a thin (6\,mm) annulus of lead around the detector or there was no shielding at all. The purpose in using several configurations was to test the robustness of the analysis and simulation under varying conditions, such as the detected rate, detector composition, background subtraction, and dead time. Table~\ref{tab:params} gives some of the rates and run parameters associated with the data set.

When irradiated with a source, the neutron rate $R_{n}$ is the difference between the total measured rate $R_{tot}$ and the uncorrelated background rate $R_{bkgd}$ that is associated with the source. The numerical values are obtained after applying both the energy and timing cuts of Section~\ref{subsec:det:fastneutron}. The neutron rate can also be expressed as

\begin{eqnarray}
R_{n} = R_{tot} - R_{bkgd} =  \epsilon A_{\mathrm{Cf}} \Omega + R_{rr} +R_{b},
\label{eqn:eff}
\end{eqnarray}

\noindent where $\epsilon$ is the detector efficiency, $A_{\mathrm{Cf}}$ is the neutron source activity, and  $\Omega$ is the solid angle subtended by the detector from the source.  $R_{rr}$ is the rate in the detector due to room-return neutrons (i.e., those source neutrons that scatter from the surrounding environment into the detector), and $R_{b}$ is the background rate in the detector when there is no source present. The measured rates must be obtained above a low-energy threshold. The values were chosen to be safely above the hardware threshold and the detector noise, and they are listed for each configuration in Table~\ref{tab:params}. 

\begin{figure}
    \centering
    \includegraphics[width=\columnwidth]{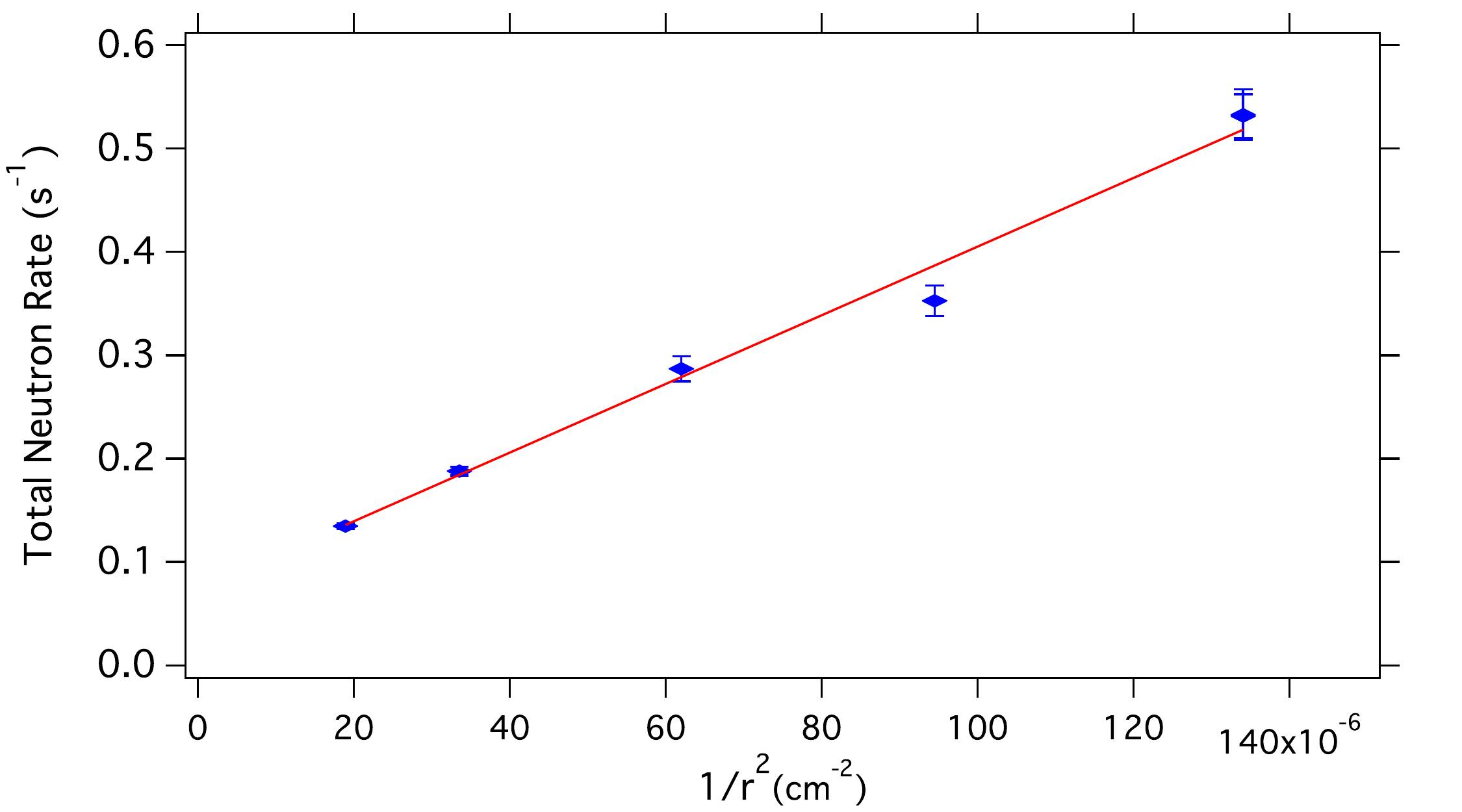}
    \caption{Plot of the total neutron rate $R_{n}$ versus $1/r^2$. The intercept of the linear fit gives $R_{rr}$, the contribution of neutrons that scatter from the room and are detected. These data are from the \Lis\ liquid scintillator with the lead annulus in place and using the higher-activity \Cftft\ source. The error bars represent plus and minus the combine standard uncertainty.}
    \label{fig:roomreturn}
\end{figure}

The background rate $R_{b}$ is negligible compared with the room return and can be ignored for these measurements. The room return contribution, however, can be a significant fraction of the measured rate and must be taken into account. The room return will be very dependent upon the geometry and material composition of the room. In the CNIF the rate is largely constant over the range of measurement positions for a given source activity~\cite{LIN83,SCH94}. To determine $R_{rr}$ for each detector and each source that were used,  measurements were taken with the source placed at several distances, $r$, from the detector. When appropriately corrected for deadtime, the intercept of a linear fit to the total rate in the detector versus $1/r^2$ gives the value of $R_{rr}$. Figure~\ref{fig:roomreturn} shows the fit for data acquired using the \Lis\ scintillator with the lead shield in place and using the higher-activity \Cftft\ source. With the measured values of the room return, the known source activity, and the calculated solid angle, one obtains the efficiency of the detector using Eq.~\ref{eqn:eff}. Table~\ref{tab:Comp} shows the results for all eight configuration under which data were acquired.

For the lifetimes, both the data and simulation were fit to a two-component exponential function, as discussed in Section~\ref{subsec:det:fastneutron}. The fast component was obtained from the fit to the simulation and held fixed in the fit to the data. For the B-plastic scintillator, the lifetime of the fast component was $0.47\,\mu$s, and for the \Lis-liquid scintillator it was equal to $1.6\,\mu$s. All the other parameters were allowed to float. This approach produced better fits to the data, particularly where the small amplitude of the component caused difficulty for the fit. The value of the long component for each configuration is given in Table~\ref{tab:Comp}. The stated uncertainties arise from statistical treatment of the data alone.

\begin{table*}
\begin{center}
\begin{tabular}{|l|cccccc|c|}
\hline
					&Pb shield	&Activity	& \multicolumn{2}{c}{Efficiency $\epsilon\ (\times 10^{-3})$}	& \multicolumn{2}{c|}{Lifetime ($\mu$s)}		\\
					&			& (Bq)	& Data 				&	Simulation				& Data 			&	Simulation		\\
\hline
 \Lis\ liquid			&	off		&	$1.2\times10^6$	&	$1.3\pm 0.1$		&	$1.4\pm 0.1$				&  $12.0\pm 2.2$	&	$10.2 \pm	0.3$		\\
scintillator				&	on		&	$1.2\times10^6$ 	&	$1.6\pm 0.1$		&	$1.7\pm 0.1$				&  $11.0\pm 1.9$	&	$10.7 \pm	0.2$		\\
					&	off		& 	$1.0\times10^3$	&	$1.3\pm 0.1$		&	$1.2\pm 0.1$				&  $14.2\pm 3.4$ 	&	$10.5 \pm	0.3$		\\
					&	on		& 	$1.0\times10^3$	&	$1.4\pm 0.1$		&	$1.4\pm 0.1$				&  $10.0\pm 2.2$ 	&	$10.8 \pm	0.4$		\\

					&			&					&					&							&				&					\\
B-plastic				&	off		&	$1.2\times10^6$	&	 	$3.0\pm 0.2$	&	$4.1\pm 0.5$				&	$1.66\pm 0.20$ 	&	$1.54 \pm	0.17$\\
scintillator				&	on		&	$1.2\times10^6$	&	 	$4.0\pm 0.2$	&	$5.0\pm 0.5$				&	$1.80\pm 0.07$ 	&	$1.69 \pm	0.06$\\
					&	off		&	$1.0\times10^3$	&	 	$3.6\pm 0.2$	&	$3.5\pm 0.5$				&	$1.78\pm 0.08$ 	&	$1.62 \pm	0.06$\\
					&	on		&	$1.0\times10^3$	&	 	$3.9\pm 0.1$	&	$4.4\pm 0.5$				&	$1.84\pm 0.07$ 	&	$1.70 \pm	0.08$\\
\hline
\end{tabular}
\end{center}
\caption{\label{tab:Comp} Comparison of the measured efficiency and capture lifetime from the MCNP model to the eight configurations of the \Lis\ liquid scintillator and B-plastic detectors. The uncertainty in the efficiency values is the standard deviation of all the individual efficiencies determined at each value of the detector-source distance $r$. The error bars for the data results come from fits to the data and represent plus and minus the combined standard uncertainty. The error bars for the simulation results represent the standard uncertainty.}
\end{table*}

\subsection{Comparison with Simulation}
\label{subsec:det:simulation}

A Monte Carlo simulation was constructed using MCNP5~\cite{BRO03} to compare with the measurements described in Section~ \ref {subsec:efficiency}.  Accurate benchmarking of this Monte Carlo model against the laboratory measurements is important for understanding the response of the detector and scintillator.  The model of the test cell consisted of a right cylinder with the same dimensions as the actual cell and filled with the 0.15\,\% \Lis-loaded liquid scintillator. The 6\,mm thick lead annulus surrounding the cylinder was also modeled to compare with the shielded measurements taken in the CNIF.  The \nuc{252}{Cf} source energy spectrum was assumed to be Maxwellian, and the stainless steel source encapsulation was modeled to take into account any attenuation arising from neutrons scattering in the source holder.  No attempt was made to model the CNIF room-return or backgrounds as they were taken into account in the measurements.

A neutron slowing down in a liquid scintillator loses its energy primarily through scattering with protons, and it is the interaction of each recoiling proton that produces the scintillation light.  If the number of scintillation photons produced is a linear function of proton recoil energy, the response of a detector can be modeled by converting the total energy deposition of a given neutron's slowing down history, a quantity easily obtainable from the standard output tally of MCNP5.  For incident gamma-rays the light-output function is linear over a wide range of energies.  For nearly all organic scintillators, however, the light-response function for heavy charged particles is non-linear. Thus, the light output from a neutron that deposits all of its energy in a single proton scatter is not the same for one that deposits all of its energy through multiple scatterings. To accurately model the detector response to fast neutrons, light output from each individual proton recoil in a given neutron history has to be summed to accurately model the scintillator response to neutrons.   We assumed that the light output of proton recoils in both the liquid \Lis-loaded scintillator and the B-loaded plastic scintillator was the same as that for the pseudocumene-based liquid scintillator NE-213 and the plastic scintillator NE-102, which is given by 
\begin{equation}
     E_{\mathrm{ee}} = 0.95 E_{\mathrm{p}} - 8.0 \left[1 - \exp(-0.1\,E_{\mathrm{p}}^{0.9}) \right],
\end{equation}
\noindent where $E_{\mathrm{p}}$ is the recoiling proton's energy, and $E_{ee}$ is the electron-equivalent light output in units of MeV~ \cite{mad78, NE80}.  

A custom post-processing routine was developed to parse the detailed neutron history output from MCNP5 (called a PTRAC file) and to extract the pertinent quantities from each of the individual scattering events for a given neutron history.  Only neutron histories in which the neutron scattered from at least one proton and was captured on a \Lis\ or \Bt\ nucleus were examined.  For each history in this subset, the electron-equivalent energy was calculated for each proton recoil and then summed to give the total light output for that particular incident neutron.  The light output was summed from the start of the neutron until the neutron was captured, and in addition, the time between the first proton scatter and the capture was recorded and histogrammed.

The comparisons between the measurements and the model results are listed in Table~\ref{tab:Comp}.  The two parameters on which we primarily focused were the efficiency and the capture lifetime. The same energy and timing cuts applied in the analysis of the data were applied to the Monte Carlo results. Ten MCNP5 runs were performed for each configuration (of source strength, shielding, and detector type), each with a different initial seed to the random number generator.  The results for detector efficiencies and lifetime were averaged across these 10 runs, and the reported error bars are the standard deviation of the mean of those values.  We did not quantify the systematic effects but believe that the two more significant effects arise as a result of uncertainty in the energy calibration of the measurements, which determines the value of the energy threshold, and knowledge of the appropriate light output function.

The agreement between the simulation and measured data is good. For the test cell containing \Lis\ liquid scintillator, an average efficiency for detecting fast \Cftft\ neutrons was on the order of $1 \times 10^{-3}$, which agreed well with the MCNP5 Monte-Carlo models.  The efficiency for the B-plastic was a few times higher, which we attribute to the larger cross section and higher concentration of the \Bt\ in the scintillator. The simulation indicates that the efficiencies for the shielded configurations were slightly higher than those of the unshielded configurations. This can be attributed to in-scattering of neutrons from the very dense lead shielding surrounding the scintillator cells.  Slight variations between the weak and strong source efficiencies can be attributed to the differences in source encapsulation.  The time distribution of the capture events also agrees well with the measured lifetimes. 

\section{Pulse-Shape Discrimination}
\label{sec:psd}

The background rejection of the scintillator was very good due primarily to the delayed capture on \Lis, as discussed in Section~\ref{subsec:intro:capturegated}. In addition to timing information, one can further discriminate between gamma rays (electronic recoil scattering) and neutrons (nuclear recoil scattering or neutron capture) based on the pulse waveform shape information. Due to the different energy deposition mechanisms, the temporal probability density functions for scintillation light creation times produced by gamma rays  and neutrons are not the same~\cite{COA08}.  Hence, one can further discriminate background gamma events from  nuclear recoil and neutron capture events~\cite{WOL95, KLE02, FLA07} based on the observed pulse produced by each event.

We describe a pulse-shape discrimination (PSD) method based on the ``Matusita distance''~\cite{MAT54, DIL78, DIL78a} between an observed pulse and nuclear and electronic recoil templates for normalized event waveforms. We also consider a standard prompt ratio method. For both methods, we estimated a discrimination threshold for accepting approximately 50\,\% of the nuclear recoil events.

\subsection{Calibration Data}\label{subsec:psdcal}

We acquired digitized pulses in two calibration measurements:  one with \Csots\ that produced essentially only gamma ray waveforms and another with a 2.5-MeV neutron generator that produced neutrons with some small admixture of gamma rays. The data acquisition system that captured the waveforms is the same as was used for the previous measurements. The purpose was to estimate the expected value of a normalized and background-corrected pulse (that sums to 1) for both the nuclear recoil and electronic recoil events. We estimated the background level as a trimmed mean of all values of the pulse. The values that contributed to the trimmed mean range from 0.125 quantile to 0.875 quantile of the distribution of observed pulse values. We registered each pulse according to the time when the pulses steeply rises from its baseline (background) plateau to 0.3 times its maximum value. For each pulse, we determined the mean value of the pulse for a 10\,ns interval before this rise time. If this mean value exceeded 0.003 times the pulse amplitude, we rejected the pulse.

\subsection{Template Estimation}\label{subsec:templ}

\begin{figure}[]
	\centering \includegraphics[width=\columnwidth]{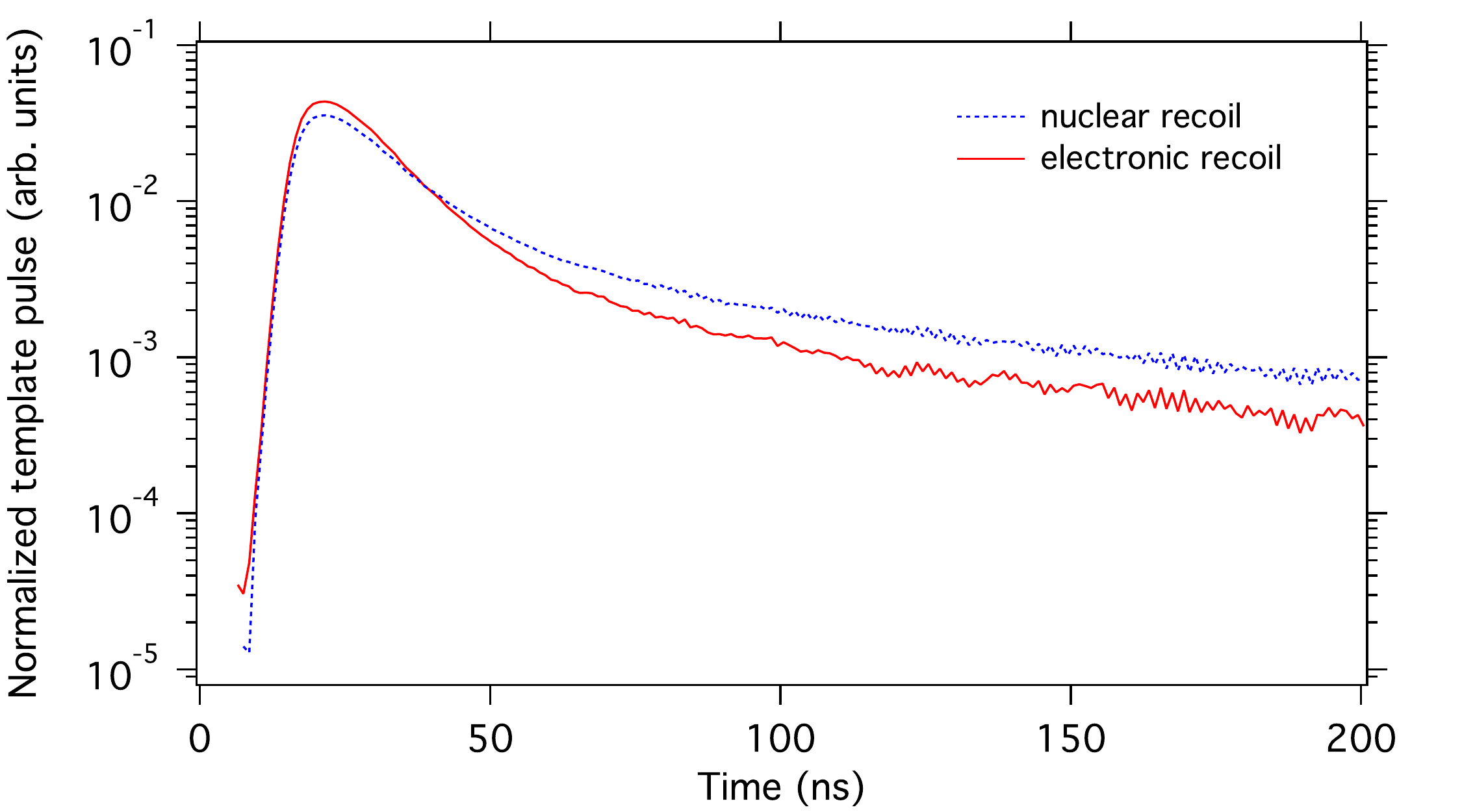}
	\caption{Waveform templates for nuclear recoil and electronic recoil events determined by cluster analysis from calibration data from a 2.5-MeV neutron source (contaminated by gamma rays)}
	\label{fig:psd1}
\end{figure}

We estimated both the nuclear and electronic recoil templates with a k-means cluster analysis~\cite{MAC67} from calibration data produced by a 2.5-MeV generator, shown in Fig.~\ref{fig:psd1}. Joint estimation of the nuclear and electronic recoil templates is possible because this calibration data is contaminated by gamma rays. Figure \ref{fig:psd1} shows the waveform templates determined from the 2.5-MeV source. We determined the templates by minimizing the squared Euclidean distance ($L^2$) of the normalized pulses within each of the two clusters. We also estimated templates with a robust version of cluster analysis based on an $L^1$ distance metric. In this approach, within each cluster, the median value rather than the mean value is computed. The robust and non-robust cluster analysis methods yield similar template estimates.

From  the ${}^{137}$Cs gamma ray source, we determined an electronic recoil template by a robust signal averaging method. Each baseline-corrected pulse was normalized so that its maximum value was 1. At each time sample, the trimmed mean of all the processed pulses was computed, and the resulting pulse was divided by its integral value. Values of the trimmed mean at each relative time of interest between the 0.1 and 0.9 quantiles of the distribution were averaged. For the 2.5-MeV source, we estimated a nuclear recoil template with the same robust signal averaging method described above. The estimated nuclear recoil and electronic recoil templates from the cluster analysis agree well with the corresponding robust signal averaging estimates. Moreover, the estimated nuclear recoil templates determined from start and stop pulses for the  2.5\,MeV case were in very close agreement for the range of amplitudes that we attribute to neutron capture on \Lis.

\subsection{Discrimination statistics}\label{subsec:discr}

The  Matusita  distances between  a normalized pulse of interest, $p_m$, and the template pulses for the electronic recoil $\hat{p}_e$ and nuclear recoil  events $\hat{p}_n$ are 

\begin{eqnarray}
d_e =  \sum_{i}^{} (~~ \sqrt{p_m(i)} - \sqrt{\hat{p}_e(i)}~~) ^2
\end{eqnarray}

and

\begin{eqnarray}
d_n =  \sum_{i}^{} (~~ \sqrt{p_m(i)} - \sqrt{\hat{p}_n(i)}~~) ^2,
\end{eqnarray}

\noindent where $i$ is the time increment for the digitized pulse. The normalized pulses sum to 1. Negative values are set to 0 before taking square roots in the above equations. Our primary PSD statistic is 

\begin{eqnarray}
\log R =  \log \frac{d_n}{d_e}.
\end{eqnarray}

\noindent For comparison, we also computed a prompt ratio statistic 

\begin{eqnarray}
f_p =  \frac{ X_p} { X_T},
\end{eqnarray}

\noindent where ${ X_p}$ is the integrated pulse from $t=0$ to $t_{o}$, and $X_T$ is the integrated pulse over all times. Here, we set $t_o$ to be the time where the nuclear and  electronic recoil pulses cross.

\begin{figure}
	\centering \includegraphics[width=\columnwidth]{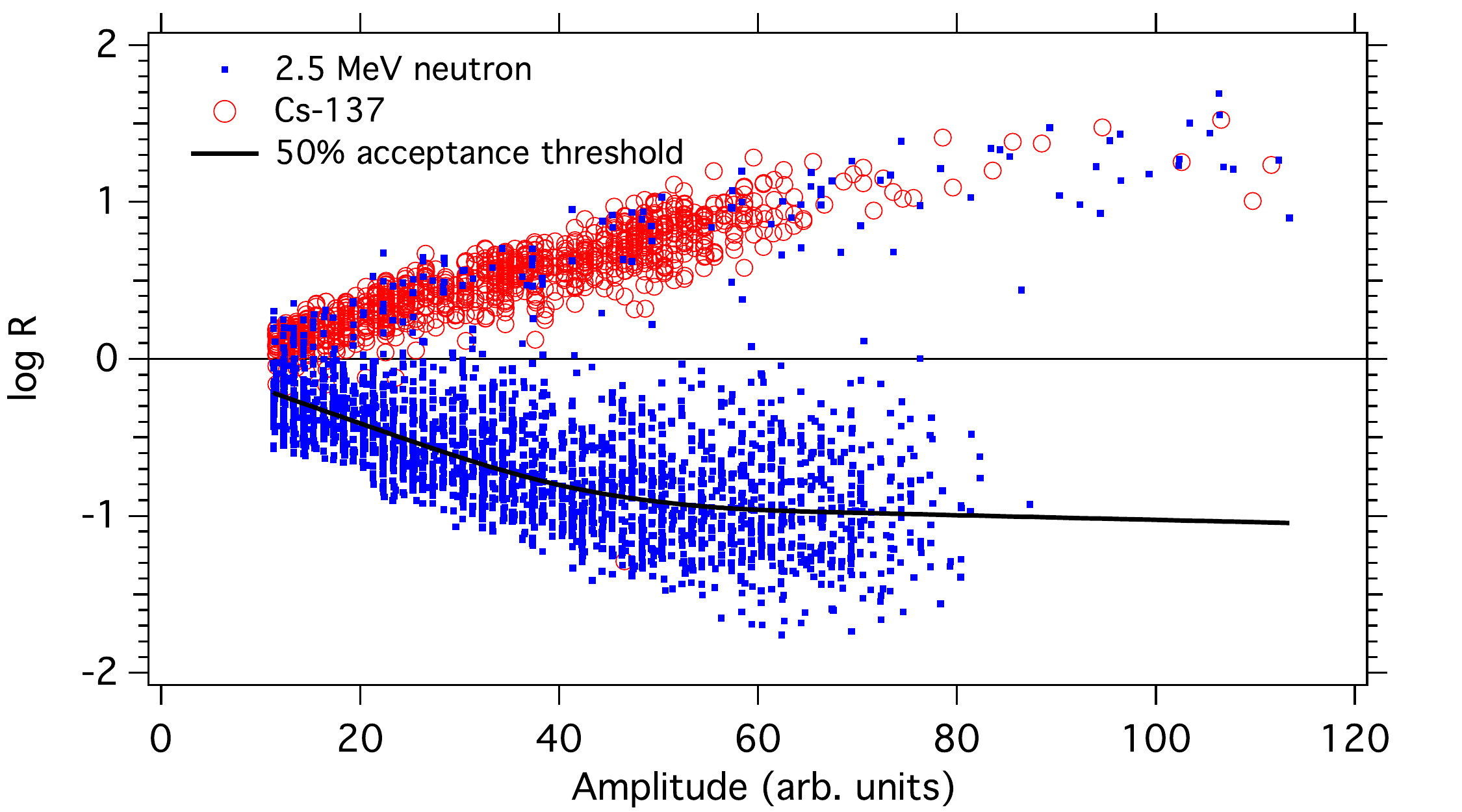}
	\caption{Empirical distribution of $\log R $ statistics.}
	\label{fig:psd2}
\end{figure}

\begin{figure}
	\centering \includegraphics[width=\columnwidth]{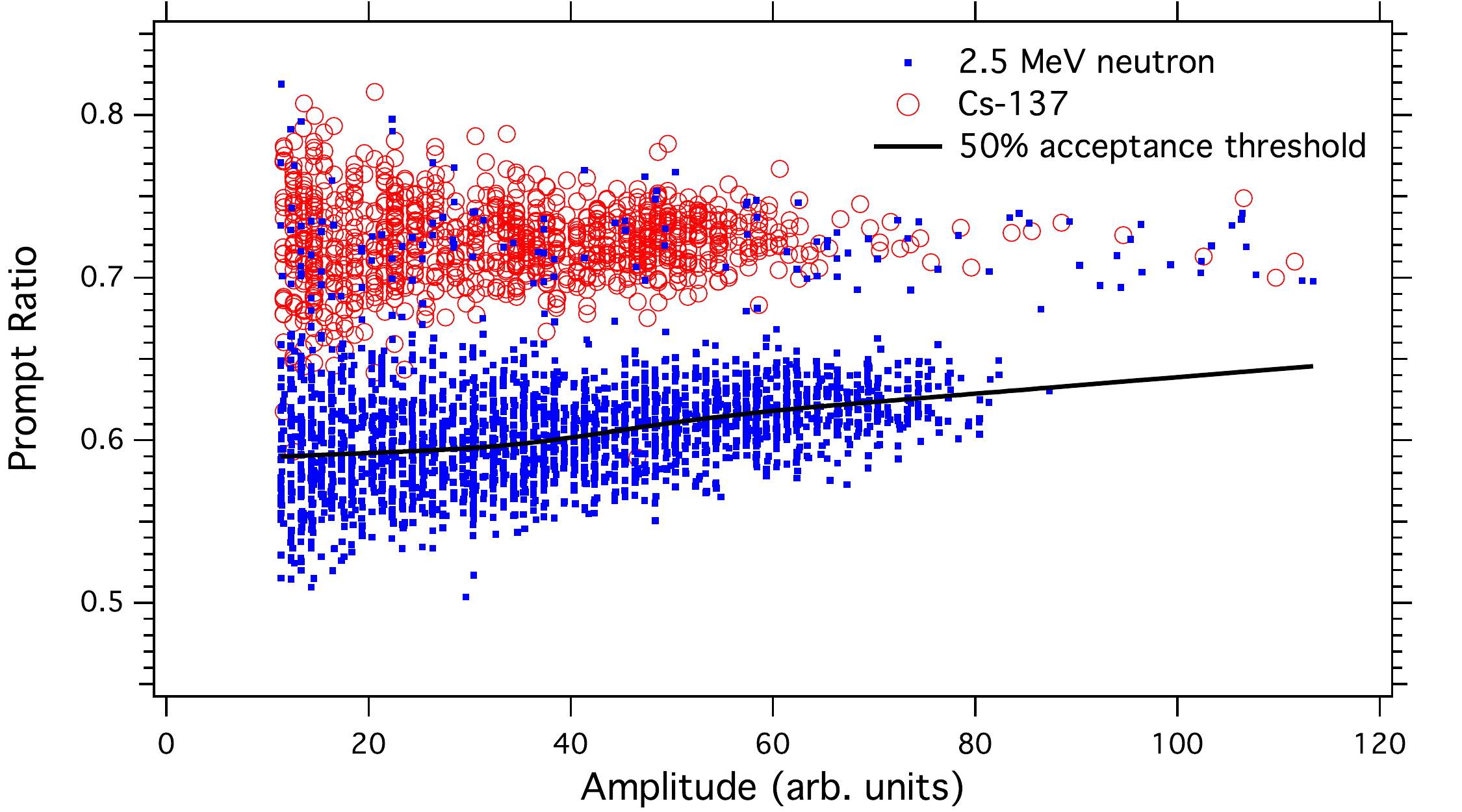}
	\caption{Empirical distribution of prompt ratio statistics. The width of the prompt time window is determined by where the nuclear and electronic recoil templates cross.}
	\label{fig:psd3}
\end{figure}

For both  discrimination statistics, Figure \ref{fig:psd2} and Figure \ref{fig:psd3}, we estimate an amplitude dependent discrimination threshold based on events that produce $\log R$ values less than 0. We then formed a curve in (amplitude, $\log R$) or (amplitude, $f_p$)  space. For each method, we sorted the corresponding curve data according to amplitude bins and determine the median amplitude and median discrimination statistic within each bin. In sequence, we fit a monotonic regression model~\cite{ROB88}  and then a smoothing spline to each curve. The degrees of freedom of the smoothing spline were determined by cross validation~\cite{HAS01} We determined a threshold for each particular amplitude by evaluating the smoothing spline model at that amplitude.

The separation between the $\log R$  statistics appears more dramatic than the separation between the $f_p$ statistics for  the ${}^{137}$Cs and 2.5-MeV sources. Theoretically, we expect that the $\log R$ statistic conveys more information because it is based on a 201-bin representation of the observed pulse whereas the prompt ratio is based on a 2-bin representation of the observed pulse. A careful quantification of the relative performance of PSD algorithms based on these
two statistics is a topic for further study. One could also form larger bins to smooth out noise before computing a $\log R$ statistic for any pulse as discussed in Ref.~\cite{LIP08,LIP10}. In future experiments, our digital acquisition system will have a higher (10-bit or 12-bit) resolution compared to the 8-bit resolution of the data shown in this study. This should facilitate refinement of our PSD techniques. In this work, we neglected to account for the energy dependence of the templates. In future work, we may account for this dependence.

\section{Summary and Conclusions}
\label{sec:conclusion}

A liquid scintillator doped with 0.15\,\% \Lis\ by weight was fabricated and made into a test cell. The process of making the scintillator does not require complicated chemical techniques, and the data acquisition system and analysis are straightforward. The cost of the raw materials was not high. The cell was tested in a known field of fast neutrons and the capture-gated detection method worked as expected. The data acquired for both the \Lis-liquid and the B-plastic scintillator were compared with Monte Carlo simulations yielding good agreement. The detection efficiency is not large (on the order of $10^{-3}$), but that is primarily due to the small size of the test cell. Starting with a detector of this volume, the efficiency scales more rapidly than the volume, so a high efficiency detector need not be prohibitively large or expensive. The spectroscopic performance of the detector was not demonstrated here, but in future work, we will investigate the response of the detector to 2.5-MeV and 14-MeV monoenergetic neutrons. 

One sample has been used in this testing over the course of about one year and did not shown noticeable degradation. There is some loss of light yield due to the introduction of the LiCl, but it is not prohibitively large. Future work will focus on quantifying optical properties and increasing the concentration of the \Lis\ in the scintillator. In principle, higher concentrations would be better in terms of increasing the efficiency and decreasing the capture time, but one also has to consider the potential degradation of light transmission in the scintillator. We note that the variety of possible \Lis-loaded scintillators has not been explored.  LiCl was chosen initially due to its relatively high solubility in water and for the ease with which LiCl can be produced from isotopically-enriched lithium carbonate.  There may be other Li-containing compounds where one can form stable emulsions with high concentrations of \Lis. Scintillators from other manufacturers should be investigated for durability and optical properties. While this paper gives data on the characteristics of detecting fast neutrons, the scintillator is naturally a very efficient detector of thermal neutrons. Future work will examine those properties.

Timing information based on the  delay time between a neutron energy deposit and its capture on \Lis\ enables one to reject backgrounds with high probability. To improve background rejection based on this timing information, we presented a  PSD method based on the log-ratio of the Matusita distances between a normalized waveform and  template waveforms for nuclear recoil and electronic recoil events. In the future, we plan to quantify the relative performance of this multidimensional method (Figure~\ref{fig:psd2}) with respect to a simpler prompt ratio discrimination method (Figure~\ref{fig:psd3}). We presented a method to determine the width of the prompt time window based on when the template waveforms cross. We also presented a nonparametric method to  determine an approximate 50 percent nuclear recoil acceptance threshold for the prompt ratio and Matusita distance PSD methods.


We thank C. Bass and C. Heimbach of NIST and E. J. Beise, H. Breuer and T. Langford of the University of Maryland for useful discussions. This work was supported in part by the American Civil Research and Development Foundation, grant no. RP2-2277.


\bibliographystyle{elsarticle-num}
\bibliography{Nico_Neutron_Ref}

\begin{thebibliography}{10}
\expandafter\ifx\csname url\endcsname\relax
  \def\url#1{\texttt{#1}}\fi
\expandafter\ifx\csname urlprefix\endcsname\relax\def\urlprefix{URL }\fi
\expandafter\ifx\csname href\endcsname\relax
  \def\href#1#2{#2} \def\path#1{#1}\fi

\bibitem{DEL95}
J.~Delorme, M.~Ericson, T.~Ericson, P.~Vogel, {Pion and neutron production by
  cosmic-ray muons underground}, Phys. Rev. D 52 (1995) 2222.

\bibitem{FOR04}
J.~Formaggio, C.~Martoff, {Backgrounds to Sensitive Experiments Underground},
  Ann. Rev. Nucl. Part. Sci. 54 (2004) 361.

\bibitem{ELL02}
S.~R. {Elliott}, P.~{Vogel}, {Double beta decay}, Ann. Rev. Nucl. Part. Sci. 52
  (2002) 115.
\newblock \href {http://arxiv.org/abs/arXiv:hep-ph/0202264}
  {\path{arXiv:arXiv:hep-ph/0202264}}.

\bibitem{AAL05}
C.~Aalseth, et~al., {The proposed Majorana Ge-76 double-beta decay experiment},
  Nucl. Instrum. Meth. B 138 (2005) 217.

\bibitem{SCH06}
S.~Sch\"onert, et~al., {Status of the Germanium detector array (GERDA) in the
  search of neutrinoless $\beta \beta$ decays of $^{76}$Ge at LNGS }, Phys.
  Atom. Nucl. 69 (2006) 2101.

\bibitem{AKE03}
D.~Akerib, et~al., {New results from the Cryogenic Dark Matter Search
  experiment}, Phys. Rev. D 68 (2003) 082002.

\bibitem{GAI04}
R.~J. {Gaitskell}, {Direct Detection of Dark Matter}, Ann. Rev. Nucl. Part.
  Sci. 54 (2004) 315.

\bibitem{BOU04}
M.~G. Boulay, A.~Hime, J.~Lidgard, {Design constraints for a WIMP dark matter
  and p p solar neutrino liquid neon scintillation detector}, nucl-ex/0410025
  (2004).

\bibitem{ANG05}
G.~Angloher, et~al., {Limit on WIMP dark matter using scintillating CaWO4
  cryogenic detectors with active background suppression}, Astropart. Phys. 23
  (2005) 325.

\bibitem{CLE98}
B.~Cleveland, et~al., {Measurement of the Solar Electron Neutrino Flux with the
  Homestake Chlorine Detector}, Astrophys. J. 496 (1998) 505.

\bibitem{HAM99}
W.~Hampel, et~al., {GALLEX solar neutrino observations: results for GALLEX IV},
  Phys. Lett. B 447 (1999) 127.

\bibitem{FUK01}
S.~Fukuda, et~al., {Solar $^8$B and hep Neutrino Measurements from 1258 Days of
  Super-Kamiokande Data}, Phys. Rev. Lett. 86 (2001) 5651.

\bibitem{ABD09}
J.~N. Abdurashitov, et~al., {Measurement of the solar neutrino capture rate
  with gallium metal. III. Results for the 2002Ð2007 data-taking period}, Phys.
  Rev. C 80 (2009) 015807.

\bibitem{MCK05}
D.~McKinsey, K.~Coakley, {Neutrino detection with CLEAN}, Astropart. Phys. 22
  (2005) 355.

\bibitem{AHA07}
B.~Aharmim, et~al., {Determination of the $\nu_e$ and total $^8$B solar
  neutrino fluxes using the Sudbury Neutrino Observatory Phase I data set},
  Phys. Rev. C 75 (2007) 045502.

\bibitem{MEI06}
D.-M. {Mei}, A.~{Hime}, {Muon-induced background study for underground
  laboratories}, Phys. Rev. D 73~(5) (2006) 053004.
\newblock \href {http://dx.doi.org/10.1103/PhysRevD.73.053004}
  {\path{doi:10.1103/PhysRevD.73.053004}}.

\bibitem{WUL04}
H.~Wulandari, J.~Jochum, W.~Rau, F.~von Feilitzsch, {Neutron Background Studies
  for the CRESST Dark Matter Experiment}, arXiv:hep-ex/0401032 v1 (2004).

\bibitem{MAR07}
M.~G. {Marino}, J.~A. {Detwiler}, R.~{Henning}, R.~A. {Johnson}, A.~G.
  {Schubert}, J.~F. {Wilkerson}, {Validation of spallation neutron production
  and propagation within Geant4}, Nucl. Instrum. Meth. A 582 (2007) 611--620.
\newblock \href {http://arxiv.org/abs/arXiv:0708.0848}
  {\path{arXiv:arXiv:0708.0848}}, \href
  {http://dx.doi.org/10.1016/j.nima.2007.08.170}
  {\path{doi:10.1016/j.nima.2007.08.170}}.

\bibitem{ISO99}
I.~O. for Standards, Reference neutron radiations: Characteristics and methods
  of production of simulated workplace neutron fields, iSO/DIS 12789.

\bibitem{KNO89}
G.~Knoll, {Radiation Detection and Measurement}, John Wiley and Sons, New York,
  1989.

\bibitem{BRO02}
F.~D. Brooks, H.~Klein, {Neutron spectrometry - historical review and present
  status}, Nucl. Instrum. Meth. A 476 (2002) 1.

\bibitem{DRA86}
D.~M. Drake, W.~C. Feldman, C.~Hurlbut, New electronically black neutron
  detectors, Nucl. Instrum. Meth. A 247 (1986) 576.

\bibitem{CZI89}
J.~B. {Czirr}, G.~L. {Jensen}, {A neutron coincidence spectrometer}, Nucl.
  Instrum. Meth. A 294 (1989) 365.

\bibitem{CZI02}
J.~B. {Czirr}, D.~B. {Merrill}, D.~{Buehler}, T.~K. {McKnight}, J.~L.
  {Carroll}, T.~{Abbott}, E.~{Wilcox}, {Capture-gated neutron spectrometry},
  Nucl. Instrum. Meth. A 476 (2002) 309.

\bibitem{NOR02}
S.~Normand, B.~Mouanda, S.~Haan, M.~Louvel, {Study of a new boron loaded
  plastic scintillator}, IEEE Trans. Nucl. Sci. 49 (2002) 1603.

\bibitem{POZ07}
S.~A. {Pozzi}, M.~{Flaska}, A.~{Enqvist}, I.~{P{\'a}zsit}, {Monte Carlo and
  analytical models of neutron detection with organic scintillation detectors},
  Nucl. Instrum. Meth. A 582 (2007) 629.
\newblock \href {http://dx.doi.org/10.1016/j.nima.2007.08.246}
  {\path{doi:10.1016/j.nima.2007.08.246}}.

\bibitem{ALE89}
R.~Aleksan, J.~Bouchez, M.~Cribier, {Measurement of fast neutrons in the Gran
  Sasso Laboratory using a $^{6}$Li doped liquid scintillator}, Nucl. Instrum.
  Meth. A 247 (1989) 203.

\bibitem{CHA98}
V.~Chazal, B.~Chambon, M.~D. Jesus, {Neutron background measurements in the
  underground laboratory of Modane}, Astropart. Phys. 9 (1998) 163.

\bibitem{AIT89}
S.~{Ait-Boubker}, M.~{Avenier}, G.~{Bagieu}, J.~F. {Cavaignac}, J.~{Collot},
  J.~{Favier}, E.~{Kajfasz}, D.~H. {Koang}, A.~{Stutz}, B.~{Vignon}, {Thermal
  neutron detection and identification in a large volume with a new lithium-6
  loaded liquid scintillator}, Nucl. Instrum. Meth. A 277 (1989) 461--466.
\newblock \href {http://dx.doi.org/10.1016/0168-9002(89)90775-4}
  {\path{doi:10.1016/0168-9002(89)90775-4}}.

\bibitem{AOY93}
T.~Aoyama, K.~Honda, C.~Mori, {Energy response of a full-energy-absorption
  neutron spectrometer using boron-loaded liquid scintillator BC-523}, Nucl.
  Instrum. Meth. A 333 (1993) 492.

\bibitem{KAL56}
H.~{Kallmann}, M.~{Furst}, F.~{Brown}, {Liquid scintillators with heavy
  elements}, Nucleonics 14 (1956) 48.

\bibitem{HEJ61}
J.~{Hejwowski}, A.~{Szymanski}, {Lithium Loaded Liquid Scintillator}, Rev. Sci.
  Instrum. 32 (1961) 1057.

\bibitem{GRE79}
L.~R. {Greenwood}, N.~R. {Chellew}, G.~A. {Zarwell}, {6Li-loaded liquid
  scintillators with pulse-shape discrimination}, Rev. Sci. Instrum. 50 (1979)
  472.

\bibitem{TRA08}
Certain trade names and company products are mentioned in the text or
  identified in illustrations in order to adequately specify the experimental
  procedure and equipment used. In no case does such identification imply
  recommendation or endorsement by the National Institute of Standards and
  Technology, nor does it imply that the products are necessarily the best
  available for the purpose.

\bibitem{FIS04}
B.~M. Fisher, J.~S. Nico, A.~K. Thompson, D.~M. Gilliam, {Development of a
  Sensitive Fast-Neutron Spectrometer}, NIST SP 1029 (2004) 50.

\bibitem{ABD02b}
J.~N. {Abdurashitov}, V.~N. {Gavrin}, A.~V. {Kalikhov}, V.~L. {Matushko}, A.~A.
  {Shikhin}, V.~E. {Yants}, O.~S. {Zaborskaia}, J.~M. {Adams}, J.~S. {Nico},
  A.~K. {Thompson}, {A high resolution, low background fast neutron
  spectrometer}, Nucl. Instrum. Meth. A 476 (2002) 318.

\bibitem{ABD07}
J.~N. {Abdurashitov}, V.~N. {Gavrin}, A.~V. {Kalikhov}, V.~L. {Matushko}, A.~A.
  {Shikhin}, V.~E. {Yants}, I.~{Veretenkin}, J.~S. {Nico}, A.~K. {Thompson}, {A
  high resolution, low background fast neutron spectrometer}, Phys. Atom. Nucl.
  70 (2007) 133.
\newblock \href {http://dx.doi.org/10.1134/S1063778807010152}
  {\path{doi:10.1134/S1063778807010152}}.

\bibitem{GRU77}
J.~Grundl, V.~Spiegel, C.~M. Eisenhauer, H.~T. Heaton, D.~M. Gilliam,
  J.~Bigelow, A californium-252 fission spectrum irradiation facility for
  neutron reaction rate measurements, Nuc. Tech. 32 (1977) 315.

\bibitem{LIN83}
L.~Linpei, Reflection of cf-252 fission neutrons from a concrete floor, Rad.
  Prot. Dosim. 5 (1983) 227.

\bibitem{SCH94}
R.~B. Schwartz, C.~M. Eisenhauer, Test of a neutron spectrometer in nist
  standard fields, Rad. Prot. Dosim. 52 (1994) 99.

\bibitem{BRO03}
F.~Brown, et~al., {A General Monte Carlo N-Particle Transport Code, Version 5},
  los Alamos LA-UR-03-1987, 2003.

\bibitem{mad78}
R.~{Madey}, F.~{Waterman}, A.~{Baldwin}, J.~{Knudson}, J.~{Carlson},
  J.~{Rapaport}, {The response of NE-228A, NE-228, NE-224, and NE-102
  scintillators to protons from 2.43 to 19.55 MeV}, Nuclear Instruments and
  Methods 151 (1978) 445.

\bibitem{NE80}
{Scintillators for the Physical Sciences}, brochure No. 126P, Nuclear
  Enterprises Inc. (1980).

\bibitem{COA08}
K.~J. Coakley, D.~F. Vecchia, J.~S. Nico, B.~M. Fisher, {Pulse Shape
  Discrimination for a Fast Neutron Detector}, {2008 IEEE Nuclear Science
  Symposium} (2008) 2984.

\bibitem{WOL95}
D.~{Wolski}, M.~{Moszy{\'n}ski}, T.~{Ludziejewski}, A.~{Johnson}, W.~{Klamra},
  {\"O}.~{Skeppstedt}, {Comparison of n-{$\gamma$} discrimination by
  zero-crossing and digital charge comparison methods}, Nucl. Instrum. Meth. A
  360 (1995) 584.

\bibitem{KLE02}
H.~{Klein}, S.~{Neumann}, {Neutron and photon spectrometry with liquid
  scintillation detectors in mixed fields}, Nucl. Instrum. Meth. A 476 (2002)
  132.

\bibitem{FLA07}
M.~{Flaska}, S.~A. {Pozzi}, {Identification of shielded neutron sources with
  the liquid scintillator BC-501A using a digital pulse shape discrimination
  method}, Nucl. Instrum. Meth. A 577 (2007) 654.
\newblock \href {http://dx.doi.org/10.1016/j.nima.2007.04.141}
  {\path{doi:10.1016/j.nima.2007.04.141}}.

\bibitem{MAT54}
K.~Matusita, {On Estimation by the Minimum Distance Metric}, Ann. Inst. Stat.
  Math. 7 (1954) 67.

\bibitem{DIL78}
W.~R. Dillon, M.~Goldstein, {On the Performance of Some Multinomial
  Classification Rules}, J. Am. Stat. Assoc. 73 (1978) 305.

\bibitem{DIL78a}
W.~R. Dillon, M.~Goldstein, {Discrete Discrimination Analysis}, John Wiley and
  Sons, New York, 1978.

\bibitem{MAC67}
J.~B. MacQueen, Some methods for classification and analysis of multivariate
  observations, in: L.~M.~L. Cam, J.~Neyman (Eds.), Proc. of the fifth Berkeley
  Symposium on Mathematical Statistics and Probability, Vol.~1, University of
  California Press, California, 1967, pp. 281--297.

\bibitem{ROB88}
T.~Robertson, F.~T. Wright, R.~L. Dykstra, Order Restricted Statistical
  Inference, John Wiley and Sons, New York, 1988.

\bibitem{HAS01}
T.~Hastie, R.~Tibshirani, J.~Friedman, The Elements of Statistical Learning,
  Springer, New York, 2001.

\bibitem{LIP08}
W.~H. Lippincott, K.~J. Coakley, D.~Gastler, A.~Hime, E.~Kearns, D.~N.
  McKinsey, J.~A. Nikkel, L.~C. Stonehill, {Scintillation time dependence in
  liquid argon}, Phys. Rev. C 78 (2008) 035801.

\bibitem{LIP10}
W.~H. Lippincott, K.~J. Coakley, D.~Gastler, A.~Hime, E.~Kearns, D.~N.
  McKinsey, J.~A. Nikkel, L.~C. Stonehill, Phys. Rev. C 81 (2010) 039901.

\end{thebibliography}
\end{document}